\documentstyle[manuscript,aps]{revtex}
\begin{document}

\preprint{IMPERIAL/TP/95-96/56}

\title{Nonperturbative calculation of symmetry breaking in quantum
field theory}

\author{Carl M. Bender\cite{bye1}\cite{bye2}}
\address{Blackett Laboratory, Imperial College, London SW7 2BZ,
United Kingdom}

\author{Kimball A. Milton\cite{bye3}}
\address{Department of Physics and Astronomy, University of Oklahoma,
Norman, OK
73019, USA}

\date{\today}
\maketitle

\begin{abstract}
A new version of the delta expansion is presented, which, unlike the
conventional delta expansion, can be used to do nonperturbative
calculations in
a self-interacting scalar quantum field theory having broken
symmetry. We
calculate the expectation value of the scalar field to first order in
$\delta$,
where $\delta$ is a measure of the degree of nonlinearity in the
interaction
term.
\end{abstract}

In a private discussion with D. Bessis we learned about the
remarkable
non-Hermitian Hamiltonian
\begin{eqnarray}
H=p^2/2+ix^3.
\label{e1}
\end{eqnarray}
The eigenvalues of this Hamiltonian are all real and positive! (We
have no proof
of this property, but there is ample numerical evidence.) This
Hamiltonian is
intriguing because it suggests a novel way to apply the {\em delta
expansion} to
quantum field theories having a broken symmetry. In this paper we
introduce our
new version of the delta expansion and use it to calculate the
(nonzero) value
of $\langle\phi\rangle$ in a scalar quantum field theory [see
Eq.~(\ref{e31})].

The delta expansion is a Taylor series in powers of $\delta$, where
$\delta$
measures the degree of {\em nonlinearity} of an interaction term
\cite{DEL1}.
For a scalar quantum field theory, the conventional
approach\cite{DEL2} has been
to introduce the parameter $\delta$ into the Euclidean Lagrangian
density by
\begin{eqnarray}
{\cal L}=(\nabla\phi)^2/2+m^2\phi^2/2+g(\phi^2)^{1+\delta}.
\label{e2}
\end{eqnarray}
The advantage of the delta expansion is that it is {\em
nonperturbative} in the
coupling constant $g$ and mass $m$ and has a nonzero radius of
convergence. The
conventional delta expansion has been used to study renormalization
\cite{DEL3},
supersymmetry \cite{DEL4}, local gauge invariance \cite{DEL5},
stochastic
quantization \cite{DEL6}, and finite-temperature field theory
\cite{DEL7}.

A drawback of the Lagrangian density in Eq.~(\ref{e2}) is that it
becomes a
$g|\phi|^3$ theory rather than a $g\phi^3$ theory when
$\delta={1\over 2}$. The
$g|\phi|^3$ theory is symmetric under $\phi\to-\phi$ and cannot
exhibit symmetry
breaking. In general, the conventional delta expansion is unsuitable
for
studying theories with symmetry breaking.

In the past it was assumed that to satisfy the physical requirement
that the
Hamiltonian be bounded below for all $\delta$, the parameter $\delta$
should
appear in the Lagrangian density (\ref{e2}) as the exponent of
$\phi^2$ and
{\em not} of $\phi$. This assumption appears to be even more
reasonable if one
expands the interaction term in Eq.~(\ref{e2}) in powers of $\delta$:
\begin{eqnarray}
g(\phi^2)^{1+\delta}=g\phi^2\sum_{k=0}^{\infty}{\delta^k\over
k!}(\log\phi^2)^k.
\label{e3}
\end{eqnarray}
One would think that the argument of the logarithm in Eq.~(\ref{e3})
should be
positive to avoid complex numbers.

However, $H$ in Eq.~(\ref{e1}) describes a theory having a positive
spectrum.
This suggests a surprising new version of the delta expansion in
which we
replace Eq.~(\ref{e2}) by
\begin{eqnarray}
{\cal L}=(\nabla\phi)^2/2+m^2\phi^2/2-g(i\phi)^{2+\delta}.
\label{e4}
\end{eqnarray}
As with Eq.~(\ref{e2}), $\delta=0$ gives free field theory. At
$\delta=1$ we
recover (when the dimension of space-time is one and the bare mass
vanishes) the
theory described by Eq.~(\ref{e1}).

We believe that for ${\rm Re}\,\delta>-2$ the quantum-mechanical
theory has a
positive real spectrum. Here, we verify this to first order in
$\delta$. The
shift in the $n$th energy level of $H=p^2/2-(ix)^{2+\delta}$ is the
expectation
value of $\delta x^2\log(ix)$ in the unperturbed harmonic-oscillator
basis
$|n\rangle$:
\begin{eqnarray}
\Delta E_n &=& \langle n| \delta x^2 \log(ix) |n\rangle \nonumber\\
&=& \delta \int_{-\infty}^{\infty} dx\, \psi_n^2(x) x^2 \log(ix)
\nonumber\\
&=& \delta\left(\int_{-\infty}^0+\int_0^{\infty}\right)
dx\,\psi_n^2(x)x^2
\log(ix)\nonumber\\
&=&\delta\int_0^{\infty}dx\,\psi_n^2(x)x^2[\log(-ix)
+\log(ix)]\nonumber\\
&=& 2\delta \int_0^{\infty} dx\, \psi_n^2(x) x^2 \log(x) \nonumber\\
&=& \delta 2^{-5/2}[a_n-(2n+1)\gamma -(5n+5/2)\log2].
\label{e5}
\end{eqnarray}
The sequence $a_n=2,\,8,\,16,\,74/3,\dots$ is given by
\begin{eqnarray}
a_{n}=4n+4+4[n/2]+(4n+2)\sum_{k=0}^{[n/2+1/2]}{1\over2k-1},
\nonumber
\end{eqnarray}
where $[\alpha]$ is the greatest integer in $\alpha$. (In the above
calculation
we choose the branch of the logarithm to lie along the negative axis
and choose
the sheet of the Riemann surface for which $\log 1=0$.) Although the
Hamiltonian
is not self-adjoint, the ${\rm O}(\delta)$ shift in the energy levels
is {\em
real\/}! This property persists to all orders in $\delta$
\cite{HMMM}.

We will now show that in any Euclidean dimension and for all
$\delta>-2$ (apart
from a special discrete set of negative values of $\delta$),
Eq.~(\ref{e4})
exhibits symmetry breaking. This result is especially interesting for
the case
$\delta=2$ because, although the resulting $\phi^4$ theory appears
symmetric
under $\phi\to-\phi$, symmetry breaking {\em persists\/}!

We begin by considering a massless {\em zero-dimensional} scalar
quantum field
theory, whose vacuum-persistence amplitude $Z$ and $N$-point Green's
function
$G_N$ are
\begin{eqnarray}
Z = \int_{-\infty}^{\infty} dx\, e^{(ix)^{2+\delta}},\quad
G_N ={1\over Z}\int_{-\infty}^{\infty} dx\, e^{(ix)^{2+\delta}} x^N.
\label{e6}
\end{eqnarray}
To evaluate $G_N$ we break the integral into two pieces:
\begin{eqnarray}
Z G_N &=& \int_{-\infty}^0 dx\, e^{(ix)^{2+\delta}} x^N +
\int_0^{\infty} dx\, e^{(ix)^{2+\delta}} x^N \nonumber\\
&=& 2 \left( {{\rm Re}~({\rm if}~N~{\rm even}) \atop i\, {\rm
Im}~({\rm if}~N~
{\rm odd})}\right)\int_0^{\infty} dx\, e^{(ix)^{2+\delta}} x^N.
\label{e7}
\end{eqnarray}
This integral exists if $-1<{\rm Re}\, \delta<1$. We let
$x^{2+\delta}=s$ and
use standard Fresnel contour integral methods to get
\begin{eqnarray}
ZG_N=-(-i)^N{2\pi\over N+1}\Bigm/{\Gamma\left(-{1+N\over
2+\delta}\right)}.
\label{e8}
\end{eqnarray}
Using this result we calculate $G_N$:
\begin{eqnarray}
G_N ={(-i)^N\over N+1}\Gamma\left(-{1\over 2+\delta}\right)
\Bigm/ \Gamma\left(-{1+N\over 2+\delta}\right).
\label{e9}
\end{eqnarray}
The nonvanishing of $G_{2N+1}$ in Eq.~(\ref{e9}) confirms that there
is
symmetry breaking.

Let us examine the case $\delta=2$ ($\phi^4$ theory) in depth. Since
the
one-point Green's function $G_1$ for this model is nonvanishing, the
symmetry
remains permanently broken as $\delta$ approaches $2$. The Green's
functions for
this theory are $G_N=(-i)^N\Gamma(3/4)/\Gamma(3/4-N/4)$. In the
presence of an
external source $J$, the vacuum functional $Z[J]$ for the theory has
a
convergent Taylor series in $J$:
\begin{eqnarray}
Z[J]=Z[0]\sum_{N=0}^{\infty}{J^N G_N\over N!}={\pi\over
2}\sum_{N=0}^{\infty}
{J^N (-i)^N\over N!\,\Gamma\left({3-N\over 4}\right)}.
\label{e10}
\end{eqnarray}
Note that $Z[J]$ obeys the third-order differential equation
\begin{eqnarray}
Z'''[J]+JZ[J]/4=0.
\label{e11}
\end{eqnarray}
This is the $D=0$ form of the functional Schwinger-Dyson equation for
the
theory. If $J$ is rotated by $e^{i\pi/4}$, Eq.~(\ref{e11}) becomes
the
differential equation satisfied by the vacuum functional $\tilde
Z[J]$ for a
conventional zero-dimensional $\phi^4$ theory in the presence of an
external
source:
\begin{eqnarray}
\tilde Z[J]=\int_{-\infty}^{\infty} dx\, e^{-x^4+Jx}.
\label{e12}
\end{eqnarray}
This model does not exhibit symmetry breaking.

The Schwinger-Dyson equation (\ref{e11}) does not specify the vacuum
functional
uniquely \cite{GG,UNIQ}. Indeed, there are four distinct solutions to
the
rotated version of Eq.~(\ref{e11}):
\begin{eqnarray}
Z_j[J] = \int_{C_j} dx\, e^{-x^4+Jx}\quad (j=1,~2,~3,~4).
\label{e13}
\end{eqnarray}
Here, $C_j$ are contours in the complex-$x$ plane; $C_1$ joins
$i\infty$ to
$+\infty$, $C_2$ joins $+\infty$ to $-i\infty$, $C_3$ joins
$-i\infty$ to
$-\infty$, and $C_4$ joins $-\infty$ to $i\infty$. (The four
solutions $Z_j[J]$
are not linearly independent; their sum is zero.) The vacuum
functional $\tilde
Z[J]$ in Eq.~(\ref{e12}) is given by $Z_4[J]+Z_1[J]$. The vacuum
functional for
the symmetry-broken theory in Eq.~(\ref{e10}) is
$e^{i\pi/4}Z_1[e^{3i\pi/4}J]$.
The symmetry-broken theory corresponding to $\delta=2$ in
Eq.~(\ref{e6}) is
distinguished from the conventional unbroken theory in
Eq.~(\ref{e12}) in that
the sign of the $x^4$ term is reversed (reflecting the factor of
$e^{3i\pi/4}$
in the source term above).

Next, let us examine the analytic behavior of Eq.~(\ref{e6}) as a
function of
complex $\delta$. The delta expansion of the one-point Green's
function $G_1$,
whose nonvanishing signals the presence of symmetry breaking, is
\begin{eqnarray}
G_1 &=& -i\sqrt{\pi}\bigm(\delta/2+\delta^2(\gamma-2-2\log
2)/8-\delta^3[48
+\pi^2 \nonumber\\
&&\quad -6(\gamma-4-2\log 2)^2]/384+\dots\bigm).
\label{e14}
\end{eqnarray}
The radius of convergence of this expansion is $2$ because there is
one
singularity in the $\delta$-plane at $\delta=-2$. [One might think
(incorrectly)
that as $\delta\to-1$ the integral for $Z$ in Eq.~(\ref{e6}) would
become a
representation of a Dirac delta function, which is singular. However,
it is not
generally true that when an integral representation for a function
ceases to
exist, the function exhibits a singularity \cite{EX}.] $G_N$ is
analytic for
${\rm Re}\,\delta>-2$ because the integration path in Eq.~(\ref{e6})
{\em is an
implicit function of} $\delta$. Indeed, as delta ranges through real
values, the
paths of integration of the two integrals in the first line of
Eq.~(\ref{e7})
{\em rotate in opposite directions} \cite{ROT}. The path of
integration of the
first integral connects $0$ and $\infty$ in the complex-$x$ plane
along the
straight line
\begin{eqnarray}
{\rm path}~1:\quad {\rm arg}\, x=3\pi/2 - \pi/(2+\delta).
\label{e15}
\end{eqnarray}
The second integration path runs from $0$ to $\infty$ along
\begin{eqnarray}
{\rm path}~2:\quad {\rm arg}\, x=\pi/(2+\delta)-\pi/2.
\label{e16}
\end{eqnarray}

When $\delta=0$ (free field theory), path 1 connects $-\infty$ to $0$
and path 2
connects $0$ to $\infty$ along the real-$x$ axis. For this case $G_N$
is
{\em real} and there is {\em no symmetry breaking}. As $\delta$
increases, path
1 rotates anticlockwise and path 2 rotates clockwise. The two paths
slope
downward at $45^\circ$ angles when $\delta=2$ ($\phi^4$ field
theory).
As $\delta\to\infty$, both paths connect the origin to $-i\infty$.
Since the
paths overlap and the integrations are in {\em opposite} senses, the
two
integrals in Eq.~(\ref{e7}) cancel. [This cancellation becomes
evident if we
set $\delta=\infty$ in Eq.~(\ref{e8}).] However, for this
$\phi^{\infty}$ theory
the expectation values $G_N$ approach finite limits:
\begin{eqnarray}
\lim_{\delta\to\infty} G_N = (-i)^N.
\label{e17}
\end{eqnarray}

The case $\delta<0$ is more interesting. As $\delta$ decreases below
$0$, path 1
rotates clockwise and path 2 rotates anticlockwise. (The paths rotate
infinitely
fast as $\delta\to -2$.) Whenever the two paths are horizontal, the
resulting
$\phi^{2/(2k+1)}$ theory is real and there is {\em no symmetry
breaking}. This
happens for the special set of values
\begin{eqnarray}
\delta=-4k/(2k+1)\quad(k=0,~1,~2,~3,\dots).
\label{e18}
\end{eqnarray}
For this case we have
\begin{eqnarray}
\lim_{\delta\to -4k/(2k+1)} G_N = \left\{\begin{array}{cc} 0 &
(N~{\rm odd}),\\
{\Gamma\left[ (N+1)({1\over 2}+k)\right]\over\Gamma\left({1\over
2}+k\right)}
& (N~{\rm even}).\end{array}\right.
\label{e19}
\end{eqnarray}
If the two paths overlap (this occurs when the paths are vertical and
go up or
down the imaginary axis), there is cancellation. This happens for a
$\phi^{1/k}$
theory when
\begin{eqnarray}
\delta=(1-2k)/k\quad(k=0,~1,~2,~3,\dots).
\label{e20}
\end{eqnarray}
For this case we have
\begin{eqnarray}
\lim_{\delta\to (1-2k)/k}G_N=(-i)^N(-1)^{kN} [k(N+1)]!/k!,
\label{e21}
\end{eqnarray}
which reduces to Eq.~(\ref{e17}) for the special case $k=0$.

Next, we examine theories in {\em one-dimensional} space-time
(quantum
mechanics). In general, the Hamiltonian
\begin{eqnarray}
H=p^2/2+(ix)^{2+\delta}
\label{e22}
\end{eqnarray}
represents a broken-symmetry theory because $H$ is not invariant
under the
parity
operation ${\cal P}$, whose effect is to make the replacements $p\to
-p$ and
$x\to -x$. (While $H$ is not symmetric under time reversal ${\cal
T}$, which
makes the replacements $p\to -p$, $x\to x$, and $i\to -i$, $H$ {\em
is}
symmetric under ${\cal PT}$.) We emphasize that there are both
broken- and
unbroken-symmetric phases in zero- and one-dimensional Euclidean
space-time, but
there is no transition between these different phases. Transitions
can only
occur when the dimension is two or more.

As in the case of zero dimensions, there is a set of values of
$\delta$ [see
Eq.~(\ref{e18})] corresponding to quantum-mechanical theories for
which there is
no broken symmetry. These special theories are described by the
Hamiltonians
\begin{eqnarray}
H=p^2/2-x^{2/(2k+1)} \quad (k=0,~1,~2,~3,\dots),
\label{e23}
\end{eqnarray}
where we have chosen $i^{2/(2k+1)}=-1$. The Hamiltonians in
Eq.~(\ref{e23})
have a deep connection with {\em quasi-exactly solvable} theories in
quantum
mechanics \cite{QES}. To express our theories in a more familiar form
we make
the change of independent variable $x=t^{2k+1}$ in the Schr\"odinger
equation
$H\psi(x)=E\psi(x)$, followed by the multiplicative change of
dependent variable
$\psi=t^k y$ to eliminate one-derivative terms. We obtain the
Schr\"odinger
equation
\begin{eqnarray}
y''(t)=\left[(k^2+k)t^{-2}-2(2k+1)^2t^{4k}(t^2+E)\right]y(t).
\label{e24}
\end{eqnarray}
Equation (\ref{e24}) represents a particle of zero energy in a
rotationally
symmetric quasi-exactly solvable potential having a centrifugal
barrier.

Finally, we use our new delta-expansion to calculate the one-point
Green's
function $G_1$ in a symmetry-broken scalar quantum field theory in
$D$-dimensional Euclidean space. The vacuum-persistence amplitude for
the
Lagrangian density in Eq.~(\ref{e4}) is
\begin{eqnarray}
Z = \int {\cal D}\phi\, e^{\int d^Dx\,\left[-{1\over 2}(\nabla\phi)^2
-{1\over 2}m^2\phi^2+g(i\phi)^{2+\delta}\right]}.
\label{e25}
\end{eqnarray}
Since the field $\phi$ has dimensions $M^{(D-2)/2}$ and the coupling
constant
$g$ has dimensions $M^{2+\delta-\delta D/2}$, where $M$ is a mass, we
re-express
Eq.~(\ref{e25}) using dimensionally explicit quantities by
substituting
$g={1\over 2}M^{2+\delta-\delta D/2}$. Expanding the exponent to
first order in
$\delta$ gives
\begin{eqnarray}
Z &=& \int{\cal D}\phi\,e^{-S_0-{1\over 2}\int d^Dx\,\left[\delta
M^2\phi^2\log
\left(i\phi M^{1-{D\over 2}}\right)+{\rm
O}(\delta^2)\right]}\nonumber\\
&=& \int {\cal D}\phi\, e^{-S_0}\Bigm[1-{1\over 2}\delta M^2\int
d^Dt\,\phi^2(t)
\nonumber\\
&&~ \times\log\left(i\phi(t)M^{1-{D\over 2}}\right)\Bigm]+{\rm
O}(\delta^2),
\label{e26}
\end{eqnarray}
where $S_0={1\over 2}\int d^D x[(\nabla\phi)^2+ \mu^2\phi^2]$ and
$\mu^2\equiv m^2+M^2$.

To evaluate the one-point Green's function $G_1$ we insert a factor
of $\phi(0)$
into the integrand of Eq.~(\ref{e26}) and divide by $Z$. By symmetry,
the
resulting expression vanishes at $\delta=0$. To first order in
$\delta$ we have
\begin{eqnarray}
G_1=-\delta{\int{\cal D}\phi\, e^{-S_0}\phi(0)\int
d^Dt\,\phi^2(t)\log\left(
i\phi(t)M^{1-{D\over 2}}\right)\over 2M^{-2}\int{\cal D}\phi\,
e^{-S_0 }}.
\nonumber
\end{eqnarray}

We now show that $G_1$ is pure imaginary: We average the above
expression with
that obtained by setting $\phi$ to $-\phi$ at every lattice point.
{\em Provided
that the integrals exist}, we obtain the explicitly imaginary result
\begin{eqnarray}
G_1=-i\pi\delta M^2{\int{\cal D}\phi\, e^{-S_0}\phi(0)\int d^Dt\,
\phi(t)|\phi(t)|\over 4\int {\cal D}\phi\, e^{-S_0}}.
\label{e27}
\end{eqnarray}

It is not easy to evaluate this functional integral because it
contains
$|\phi(t)|$, so we employ a limiting process: $\lim_{\alpha\to
1/2}\phi^{2
\alpha}=|\phi|$. Before taking the limit we temporarily treat
$\alpha$ as an
integer. Thus, this limiting process involves continuing off the
integers. [We
cannot prove the validity of this technique, but we justify this
procedure in
part by comparing the final result with the first term in
Eq.~(\ref{e14}) for
the case $D=0$.] To evaluate this limit we introduce an external
source $J(x)$,
and express $G_1$ in terms of functional derivatives with respect to
$J(x)$:
\begin{eqnarray}
G_1 &=& -i{\pi\over 4}\delta M^2 \int d^Dt\,{\delta\over\delta
J(0)}\nonumber\\
&&\quad\times\lim_{\alpha\to 1/2}\left({\delta\over\delta
J(t)}\right)^{1+2
\alpha}{{\cal Z}[J]\over {\cal Z}[0]} \Bigg|_{J=0},
\label{e28}
\end{eqnarray}
where
\begin{eqnarray}
{\cal Z}[J] &\equiv& \int{\cal D}\phi\, e^{\int d^Dx\,\left[-{1\over
2}
(\nabla\phi)^2-{1\over 2}\mu^2\phi^2 +J\phi\right]}\nonumber\\
&=& {\cal Z}[0] e^{{1\over 2}\int\int d^Dx\,d^Dy\,
J(x)J(y)\Delta(x,y)},
\label{e29}
\end{eqnarray}
and $\Delta(x,y)$, the free propagator in $D$-dimensional Euclidean
space, is
the Fourier transform of $1/(p^2+\mu^2)$:
\begin{eqnarray}
\Delta(x,y) &=& \int{d^Dp\over(2\pi)^D}e^{-ip\cdot(x-y)}{1\over
p^2+\mu^2}
\nonumber\\
&=&{1\over 2\pi}\left({2\pi|x-y|\over\mu}\right)^{1-{D\over
2}}K_{-1+{D\over 2}}
(\mu|x-y|).
\label{e30}
\end{eqnarray}

For integer $\alpha$ we use Eqs.~(\ref{e29}) and (\ref{e30}) to
evaluate
Eq.~(\ref{e28}) and obtain the final result for $G_1$:
\begin{eqnarray}
G_1 &=&-i{\pi\over 4}\delta M^2 \lim_{\alpha\to 1/2} \int d^Dt\,
{\delta\over
\delta J(0)}\left({\delta\over\delta
J(t)}\right)^{1+2\alpha}\nonumber\\
&&\times\left[{1\over 2}\int\int d^Dx\,d^Dy\,J(x)J(y)\Delta(x,y)
\right]^{\alpha+1} /(\alpha+1)!\nonumber\\
&=&-i{\pi\over 4}\delta M^2 \lim_{\alpha\to 1/2}
{(2\alpha+2)!\Delta^{\alpha}(0,0)\int d^Dt\, \Delta(0,t) \over
2^{\alpha+1} (\alpha+1)! }\nonumber\\
&=&-i\delta M^2\sqrt{\pi\Delta(0,0)/2}\int
d^Dt\,\Delta(0,t)\nonumber\\
&=&-i\delta M^2\mu^{-3}(\mu^2/4\pi)^{D/4}\sqrt{\pi\Gamma(1-D/2)/2},
\label{e31}
\end{eqnarray}
where we have used $\int d^Dt\,\Delta(0,t)=\mu^{-2}$ and
\begin{eqnarray}
\Delta(0,0)=\mu^{D-2}2^{-D}\pi^{-D/2}\Gamma(1-D/2)\quad (D<2).
\label{e32}
\end{eqnarray}
At $D=0$, $m=0$, and $M=\mu=\sqrt{2}$, Eq.~(\ref{e31}) reduces to the
first
term in Eq.~(\ref{e14}). Note that as $D$ increases past $2$, $G_1$
in
Eq.~(\ref{e31}) becomes {\em real} because the gamma function changes
sign. For
$D<2$, $\Delta(0,0)$ is finite and positive, but for $D>2$ it is a
divergent
integral regulated by continuing through complex dimension. Thus, for
$D>2$, the
sign is {\em a priori} unpredictable.

We thank D.~Bessis, G.~Guralnik, and Z.~Guralnik for their invaluable
insight.
CMB is grateful to the Physics Department at the Technion -- Israel
Institute of
Technology and the Theoretical Physics Group at Imperial College,
London, for
their hospitality and he thanks the Lady Davis Foundation, the
Fulbright
Foundation, and the PPARC for financial support. CMB and KAM thank
the
U.S.~Department of Energy for financial support.


\begin{references}
\bibitem{HMMM} To order $\delta^2$ the energy remains real because
the imaginary
parts in $\langle n|H_1|m\rangle\,\langle m|H_1|n\rangle$ cancel as
in
Eq.~(\ref{e5}). We are convinced that similar cancellations occur to
all orders.
\bibitem{EX} A simple example is the integral $\int_0^{\infty} dt\,
e^{-t/z}$,
which exists for ${\rm Re}\,z>0$. When this condition is met, the
integral
evaluates to $z$, which is an {\em entire} function of $z$.
\end{references}
\end{document}